\documentclass[amsmath, amssymb, preprintnumbers, showpacs, showkeys,aps,prl,superscriptaddress,twocolumn]{revtex4-2}

\usepackage{amsmath}
\usepackage{amssymb}
\usepackage{gensymb}
\usepackage{bm}
\usepackage{graphicx}
\usepackage{float}
\usepackage[dvipsnames]{xcolor}
\usepackage{placeins}
\usepackage{braket}
\usepackage{bbold}
\usepackage{ulem}
\usepackage[colorlinks, linkcolor=blue, citecolor=blue, urlcolor=blue, breaklinks=red]{hyperref}
\usepackage{tikz}
\usetikzlibrary{shadings}

\newcommand{\ketbra}[2]{|#1\rangle\langle #2|}

\newcommand{\mi}{ {\rm i} }
\newcommand{\me}{ {\rm e} }
\usepackage{bbm}
\newcommand{\1}{\mathbbm{1}}

\usepackage{dsfont}
\usepackage{orcidlink}

\newcommand{\UFSCar}{Departamento de Física, Universidade Federal de São Carlos, \\Rodovia Washington Luís, km 235 - SP-310, 13565-905 São Carlos, SP, Brazil}
\newcommand{\CESQ}{CESQ/ISIS (UMR 7006), CNRS and Universit\'{e} de Strasbourg, 67000 Strasbourg, France}

\definecolor{Myblue}{HTML}{648fff}
\definecolor{Myred}{HTML}{cc6677}
\begin{document}
 
\title{Postselected Entangled States by Photon Detection}

\author{P. Rosario~\orcidlink{0000-0002-7628-7373}}
\affiliation{\UFSCar}
\affiliation{\CESQ}

\author{A. Cidrim~\orcidlink{0000-0003-0007-2330}}
\affiliation{\UFSCar}

\author{R. Bachelard~\orcidlink{0000-0002-6026-509X}}
\affiliation{\UFSCar}

\begin{abstract}
Postselection is a non-deterministic mechanism to entangle subsystems, often used in weakly-excited systems. We here show how highly-excited ensembles of two-level emitters can be entangled by photon detection. A collective spin is formed, characterized by a squeezing parameter detected by far-field measurements. While decoherence is detrimental to this conditional entanglement, successive photon detections act as a purification process and restores the spin squeezing. Our work opens up new avenues for the generation of postselected entanglement in open quantum systems.
\end{abstract}
	
\date{\today}

\maketitle



{\it Introduction---}The generation and characterization of entanglement is one of the main goals of quantum physics research. Its scalability with the number of constituents of a system is a challenging but central task for both fundamental and practical use in quantum information technologies \cite{Horodecki_2009}. In physical systems such as trapped ions \cite{Bruzewicz_2019,Northup_2022} or cold atoms \cite{Takahashi_2009,Mitchell_2014}, entanglement via spin squeezing is often generated dynamically, generally using one- \cite{K_U_1993,Oberthaler_2008,Polzik_2008,Nayak_2006} or two-axis twisting Hamiltonians \cite{Helmerson_2001,Klaus_2002,Lukin_2002}.

Other successful approaches to produce entanglement involve postselection of states, harnessing projective measurements. Such methods have allowed entangling physically separate ensembles of atoms \cite{Cabrillo_1999,Monroe_2007,Blatt_2013,2022_Luo_Pan_prl}, opening the possibility to the creation of quantum networks, which aim to transmit information efficiently over large distances~\cite{Duan_2001,Zhou_2022,2025_Hanni_deRiedmatten_prx}. In this sense, postselection has proven to be a scalable method, with the experimental demonstration of large numbers of particles entangled~\cite{Kimble_2006}. 

Nevertheless, although photon detection had its merit demonstrated to serve as a herald for entanglement, many questions remain open in the field. Indeed, postselection protocols to entangle large atomic systems have been extensively applied in a weak drive (almost single-photon) regime in the context of the seminal DLCZ protocol~\cite{Duan_2001} for quantum memories \cite{Riedmatten_2021,Wei_2020}. Yet these schemes exploit multilevel internal structures and Raman scattering~\cite{2007_Laurat_Kimble_prl, 2021_Li_Jin_opt} to generate heralded spin-wave excitations from ensembles initially prepared in their electronic ground state, often in optical cavities~\cite{2012_Ritter_Rempe_nat,2015_Reiserer_Rempe_rmp}. A question which remains open is whether ensembles of two-level emitters in free space, with many electronic excitations, can be used, with spontaneously emitted photons heralding the entanglement of spins on the same transition.


In this Letter, we show how postselected entanglement can be generated from large systems of two-level atoms with arbitrary number of excitations. We introduce a spin squeezing parameter to characterize the entanglement, which can be accessed by far-field measurements of the electric field fluctuations. Applicable to pure states, we show that mixed states are also good candidates for this heralding process. Indeed, successive photon detections act as a purification process, allowing for the rise of the spin squeezing. Our work thus opens up new avenues for the generation of postselected entangled states in large quantum systems prone to decoherence. 
\begin{figure}[b!]
\centering
\includegraphics[width=\columnwidth]{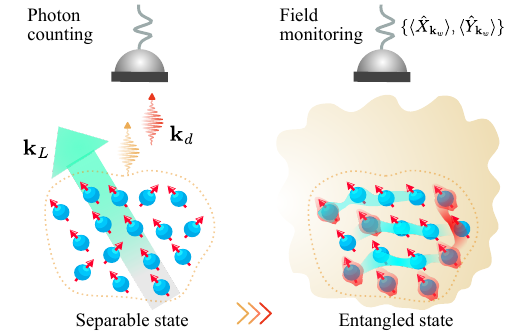}
\caption{
An ensemble of atoms initially in a separable state is becomes entangled after several photon detections in direction $\mathbf{k}_{d}$. The drive $\mathbf{k}_{L}$ is used to create the (separable) initial state. The directions in which the photons are detected, $\mathbf{k}_{d}$, and the electric field measured, $\mathbf{k}_{w}$, are correlated with the direction of the created collective spin.
}
\label{Fig:1}
\end{figure}

{\it Projection by photon detection---}In order to focus on the generation of entanglement by detection, we hereafter consider non-interacting systems prepared in separable/product states, so no entanglement is present in the initial state. The initial ensemble of $N$ spin-$1/2$ particles is described by a density matrix $\hat{\rho}_{0}=\bigotimes_{p=1}^{N}\hat{\rho}_{p}$, with $\hat{\rho}_{p}$ the single-spin state. 
In the far field, the detection of a photon along the direction $\mathbf{k}_{d}\equiv k\hat n$ (with $k$ the wavenumber of the atomic transition and $\hat n$ an observation direction) scattered from the atomic ensemble is described by the positive part of the electric field~\cite{Agarwal1974}
\begin{align}
\hat{E}_{\mathbf{k}_{d}}^{+}=\sum_{p=1}^{N}\me^{-\mi\mathbf{k}_{d}.\mathbf{r}_p}\hat{\sigma}_{p}^{-},\label{eq:E}
\end{align}
with $\hat{\sigma}^{-}_p=\ketbra{g}{e}_p$ the single-spin lowering operator between the ground and excited states $\ket{g}$ and $\ket{e}$ of the $p$-th atom, and $\mathbf{r}_p$ its position. 

The far-field operator~\eqref{eq:E} is collective, i.e., the spins are indistinguishable through the detection, which allows for the generation of postselected entanglement (see sketch in Fig.~\ref{Fig:1}). 
The quantum state after detecting $\nu$ photons along the same direction $\mathbf{k}_{d}$ is given by
\begin{align}
\hat{\rho}_{\nu}=\frac{\left(\hat{E}^{+}_{\mathbf{k}_{d}}\right)^{\nu}\hat{\rho}_{0}\left(\hat{E}^{-}_{\mathbf{k}_{d}}\right)^{\nu}}{\text{Tr}\left[\left(\hat{E}^{+}_{\mathbf{k}_{d}}\right)^{\nu}\left(\hat{E}^{-}_{\mathbf{k}_{d}}\right)^{\nu}\right]}.
\label{eq:detected_photons}
\end{align}


To test for possible entanglement in $\hat{\rho}_{\nu}$ after these $\nu$ photon detections, we use field-based entanglement witnesses~\cite{Rosario_2024}. This approach has the advantage to apply to both pure and mixed states, and rely only on macroscopic measurements of the field quadratures, $\hat{X}_{\mathbf{k}_{w}}=\hat{E}^{+}_{\mathbf{k}_{w}}+\hat{E}^{-}_{\mathbf{k}_{w}}$, $\hat{Y}_{\mathbf{k}_{w}}=\mi(\hat{E}^{+}_{\mathbf{k}_{w}}-\hat{E}^{-}_{\mathbf{k}_{w}})$ along a direction $\mathbf{k}_{w}$, and of the inversion population $\hat{Z}=\sum_{p=1}^{N}\hat{\sigma}^{z}_{p}$ (with $\hat{\sigma}^{z}=\ketbra{e}{e}-\ketbra{g}{g}$). Given these definitions, we now introduce the field-based squeezing parameter
\begin{equation}
    \xi^{2} =\frac{\text{min}_{\hat{\chi}\in\{\hat{X}_{\mathbf{k}_{w}},\hat{Y}_{\mathbf{k}_{w}},\hat{Z}\}}\left((N-1)(\Delta \hat{\chi} )^{2}+\langle \hat{\chi}^{2}\rangle\right)}{\langle \hat{E}^{2}\rangle - 2N},
    \label{eq:xi_parameter}
\end{equation}
with $(\Delta \bullet )^{2}=\langle \bullet ^{2}\rangle - \langle \bullet \rangle ^{2}$ the operator variance and  $\hat{E}^{2}=\hat{X}_{\mathbf{k}_{w}}^{2}+\hat{Y}_{\mathbf{k}_{w}}^{2}+\hat{Z}^{2}$ the length of operators $(\hat{X}_{\mathbf{k}_{w}},\hat{Y}_{\mathbf{k}_{w}},\hat{Z}$) -- see~\cite{SM} for its derivation. Any quantum state $\hat{\rho}$ which satisfies $0\leq \xi^{2}<1$ is entangled.


In the particular case where all phase terms $\mathbf{k}_{w}\cdot\mathbf{r}_{p}$ are all equal, such as for a one- or two-dimensional chain of emitters monitored in a transverse direction, $\hat{X}_{\mathbf{k}_{w}}^{2}$ and $\hat{Y}_{\mathbf{k}_{w}}^{2}$ reduce to the usual collective spin operators, and $\hat{E}^{2}$ to the spin length. With the extra hypothesis of a maximum length, $\langle \hat{E}^{2}\rangle=N(N+2)$, Eq.~\eqref{eq:xi_parameter} reduces to the Kitagawa-Ueda spin squeezing parameter~\cite{K_U_1993,Nori_2011}. In this sense, the $\xi^2$ in Eq.~\eqref{eq:xi_parameter} corresponds to a generalized spin squeezing parameter, which encapsulates the phase information on the direction of observation. 

{\it Conditional entanglement of initial pure states---}Let us first illustrate the generation of postselected entanglement by photon detection from coherent spin states (CSSs)~\cite{Gilmore_1990}. These states are at the heart of Ramsey protocols and have been used to create multipartite entanglement through controlled squeezing operations~\cite{Kitagawa_2001}. In the generalized Bloch sphere formalism, a CSS is represented by a collective spin, whose state reads
\begin{align}
\label{eq:pure}    \hat{\rho}_{0}=\bigotimes_{p=1}^{N}\begin{pmatrix}
\sin^{2}\left(\theta/2\right) & \me^{\mi\phi_p}\sin\left(\theta\right)/2 \\
\me^{-\mi\phi_p}\sin\left(\theta\right)/2 & \cos^{2}\left(\theta/2\right)\\
\end{pmatrix},
\end{align}
parametrized by a polar angle $\theta$ and a phase $\phi_p=\mathbf{k}_{L}.\mathbf{r}_{p}$ imprinted by a driving laser.

For a fully-excited cloud of $N$ atoms ($\theta=\pi$, $\hat{\rho}_{0}=\ketbra{e}{e}^{\otimes N}$), detecting $\nu\leq N-1$ photons along a direction $\mathbf{k}_{d}$ and monitoring the squeezing parameter along the same direction ($\mathbf{k}_{w}=\mathbf{k}_{d}$) leads to~\cite{SM}
\begin{align}
     \xi^{2}=\frac{(N-2\nu)^{2}}{N^{2}}.
     \label{eq:xi_full_exc}
\end{align}
In this case, the squeezing is independent of the chosen direction for detecting and measuring the squeezing since the optical phase of each atom present in Eqs.~\eqref{eq:E} and in the quadratures can be absorbed by renormalization of the excited state.

This situation is actually the one envisioned by Dicke for the superradiant cascade from a fully excited state~\cite{Dicke_1954}, where the system passes through a series of entangled states, today known as the Dicke basis. The strongest squeezing in Eq.~\eqref{eq:xi_full_exc} is obtained when $\nu= N/2$, with $\xi^2=0$ for $N$ even and $1/N^2$ for $N$ odd, which also corresponds to the maximum emission rate of the  cascade. Note that the present generation of entanglement relies on the (postselected) detection of multiple photons in a single experimental realization (trajectory), which is equivalent to monitoring higher-order (auto)correlations functions $g^{(m)}$ at zero delay~\cite{Zanthier_2006,Bojer_2022,Wolf_2020} for each experimental run. It is important to note that this single-trajectory measurement, involving multiple photon detections within one experimental run, is fundamentally different from the ensemble-averaged evolution captured by a master equation, where repeated measurements of the initially fully excited system average out to a non-entangled mixed state.~\cite{Yelin_2014,bassler_2025,Zhang_2025,Rosario_2025}. This difference highlights the special role of detection in quantum mechanics~\cite{Zhang_2017}, and the associated projective nature of the measurement. 

\begin{figure}[b!]
\centering
\includegraphics[width=\columnwidth]{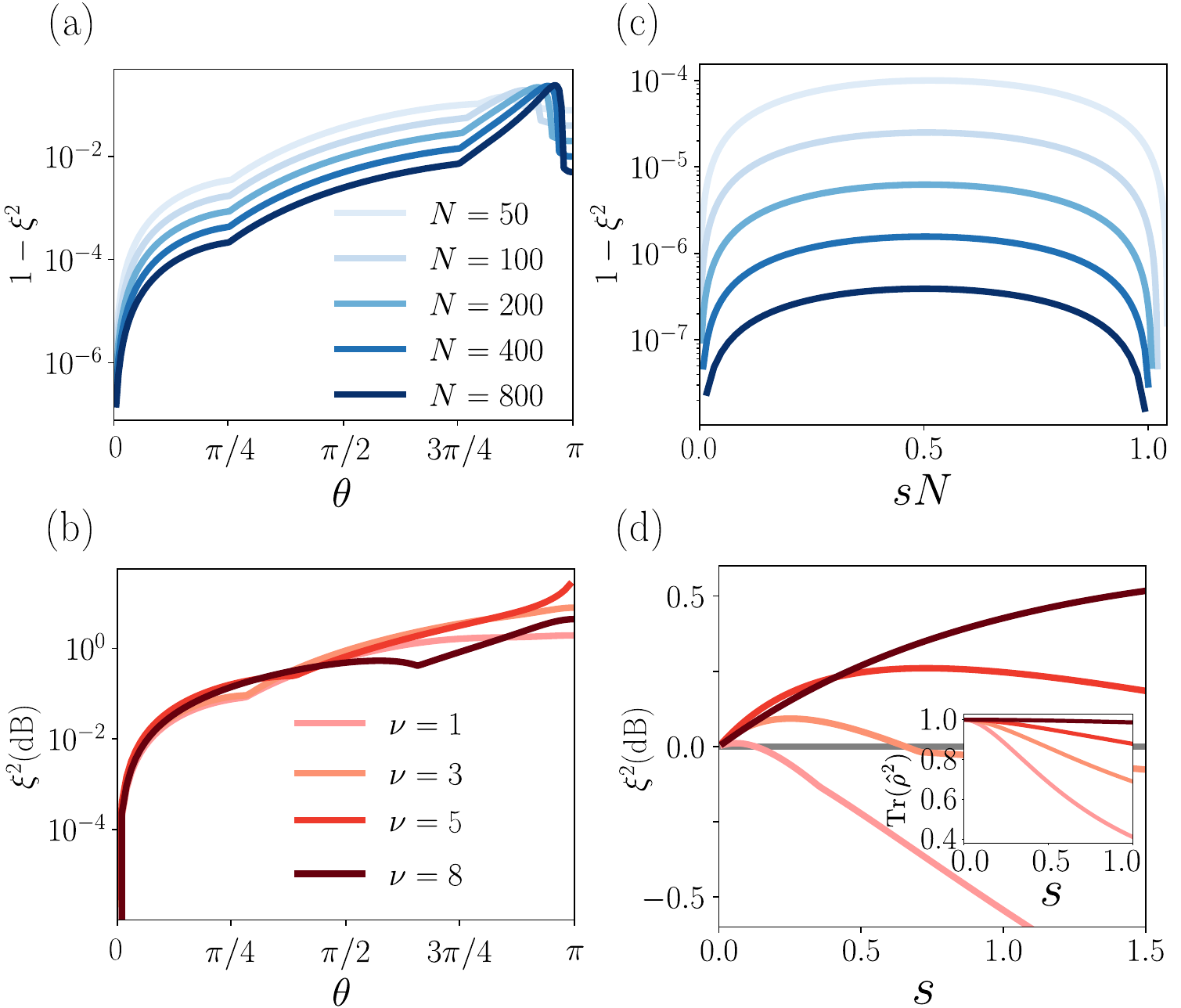}
\caption{
Squeezing parameter $\xi^2$ produced in a regular chain along $\hat{z}$-axis and with a lattice step $d=2\pi/k$ for a system initially (a-b) in a CSS~\eqref{eq:pure}, with the drive, the photon and squeezing detection all along the $x$-axis; (c-d) in a steady state~\eqref{eq:driven_state} reached with a laser at an angle $\theta_{L}=\pi/3$ from the $\hat{z}$-axis, with the photon and squeezing detection along the same direction. In (a) and (c), $\xi^2$ is computed after detecting a single photon ($\nu=1$) from a chain of $N=50$ to $800$ emitters; for (b) and (d), $\nu=1$ to $8$ photons detected from a chain of $N=10$ emitters. The inset in (d) illustrates the purification process occurring with the detection events. 
}
\label{Fig:2}
\end{figure}

Moving away from Dicke's scenario, we now consider the more general case of arbitrary CSSs~\eqref{eq:pure}. For $\theta<\pi$, despite no longer exploring Dicke states, the detection of a single photon is already sufficient to generate an entangled state, see Fig.~\ref{Fig:2}(a). While the squeezing created is larger for CSS close to $\theta=\pi$, interestingly the fully excited state does not generate the largest value, but rather values slightly below $\pi$. In other words, the fully excited state, which is the initial state considered for the Dicke cascade to decay through entangled (Dicke) states by successive quantum jumps (a picture which sparkled the debate on the entangled nature of the process~\cite{Harry_1972,Yelin_2014,Zhang_2025,Rosario_2025}) is not the state that generates the strongest squeezing from a single photon detection.

However, the detection of multiple photons changes this scenario as it enhances the generated squeezing, as shown in Fig.~\ref{Fig:2}(b) [$\xi^{2}(\text{dB})=-10\log_{10}(\xi^{2})$]: For $N=10$, the maximum squeezing is reached for $\nu=N/2=5$ detected photons, and for the initially fully inverted system, $\theta=\pi$. In this sense, the Dicke cascade remains the best candidate to generate strongly squeezed states by multiple quantum jumps (that is, photon detections). 

 
{\it Laser-driven steady state---}Let us now leave the Hilbert space and consider initial mixed states. The atoms are driven by a coherent plane-wave, with wavevector $\mathbf{k}_{L}$ and saturation parameter of $s$, until steady state. Neglecting direct interactions between the emitters, the resulting separable state reads
\begin{align}
\hat{\rho}_{0}=\bigotimes_{p=1}^{N}\begin{pmatrix}
\frac{s}{2(1+s)} & \mi\me^{\mi\mathbf{k}_{L}\cdot \mathbf{r}_{p}}\sqrt{\frac{s}{2(1+s)^{2}}}\\
-\mi\me^{-\mi\mathbf{k}_{L}\cdot \mathbf{r}_{p}}\sqrt{\frac{s}{2(1+s)^{2}}} & \frac{2+s}{2(1+s)}\\
\end{pmatrix}.
\label{eq:driven_state}
\end{align}
Then, for a fully mixed state ($\hat{\rho}_{0}=\frac{1}{2^{N}}\bigotimes_{p=1}^{N}\1_{p  }^{2\times 2}$, which is reached for $s\to\infty$), after detecting $\nu\leq N-1$ photons in direction $\mathbf{k}_{d}$, the squeezing parameter along $\mathbf{k}_{w}$ reads~\cite{SM}
 \begin{align}
     \xi^{2} =\frac{\nu^{2}+N(N-\nu)}{N+\nu(\nu-1)+\frac{2\nu(N-\nu)}{N(N-1)}f(\mathbf{k}_{d}-\mathbf{k}_{w})},
     \label{xi_remark_2}
 \end{align}
with $f(\mathbf{k}_{d}-\mathbf{k}_{w})=\left|\sum_{p=1}^{N}\me^{-\mi(\mathbf{k}_{d}-\mathbf{k}_{w}).\mathbf{r}_{p}}\right|^{2}-N$ a structure factor. Eq.~\eqref{xi_remark_2} is maximized for $\mathbf{k}_{d}=\mathbf{k}_{w}$ (a condition which we discuss later), and the squeezing can be shown to arise ($\xi^2<1$) once $\nu= \text{floor}[(N+1)/2]$ photons are detected~\cite{SM}. In the large $N$ limit, it reaches a maximum of $\xi^2=\sqrt{3}/2$ for $\nu=\text{floor}[(\sqrt{3}-1)N]$. Hence, even a fully mixed state can be turned into a squeezed state by photon detection, although the resulting squeezing is ultimately bounded by $\xi\approx 0.62$dB, much less than in the case of CSSs.


Indeed, increasing the number of excitations of a laser-driven steady-state comes with a lower purity, as spontaneous emission (a form of decoherence) starts to manifest. As a consequence, steady states with a largest excited population are not necessarily optimal to generate squeezing by photon detection: This is illustrated in Fig.~\ref{Fig:2}(c) where, for a given $N$, one observes that the saturation parameter providing the maximum squeezing after detecting a single photon is $s=1/2N$. Furthermore, the limit $sN\to 1$ shows that $\xi^2$ goes above unity (absence of squeezing) if the number of excitations in the system (which is equal to $sN$ when $s\ll 1$~\cite{Steck_2025_notes}) is larger than one. This is an important difference with CSSs, for which larger numbers of excitations lead to stronger squeezing, already conditioned to a single detected photon. It is thus a signature on the dramatic effect of decoherence on the possibility to generate conditional entanglement.

One way to overcome this issue is to detect successive photons: As depicted in Fig.~\ref{Fig:2}(d) for a regular chain of $N=10$ emitters and $v=1,3,5,8$ detected photons, the larger the number of detected photons, the stronger the squeezing and the larger the range of $s$ in which it can be detected. In particular, from $\nu=N/2=5$ detected photons, any value of $s$ leads to a squeezed state, which is consistent with result~\eqref{eq:xi_full_exc} for the fully mixed state. The role of decoherence in the conditional generation of entanglement can be appreciated in the inset of Fig.~\ref{Fig:2}(d): While the purity $\text{Tr}(\hat{\rho}^2)$ decreases with the saturation parameter $s$, increasing the number of detected photons corresponds to a purification process, as shown by the rise of purity with $\nu$. 
This can be interpreted as the fact that detecting spontaneously emitted photons reduces the information lost to the environment, thus restoring the purity to the state.



{\it Population states---}Considering the detrimental role of decoherence and the importance of (optical) coherence for the steady states discussed above, let us now investigate the possibility to conditionally generate squeezing from population states, such as those obtained by driving incoherently quantum emitters or through two-photon transitions. Considering all atoms driven equally, the associated density matrix is
\begin{align}
    \hat{\rho}_{0}=\frac{1}{2^{N}}\bigotimes_{p=1}^{N}\begin{pmatrix}
\sin^2(\overline{\theta}/2) & 0 \\
0 & \cos^2(\overline{\theta}/2)\\
\end{pmatrix},
\label{eq:no_coherence}
\end{align}
so $\sin^2(\overline{\theta}/2)$ stands for the excited population, similarly to the CSSs.
Then, assuming $\mathbf{k}_{d}=\mathbf{k}_{w}$, 
the squeezing parameter after detecting $1\leq \nu < N-1$ photons scattered from $N>2$ atoms is given by~\cite{SM}
\begin{align}
    \xi^{2}= \frac{N^{2}+(\nu^{2}-\nu N)(1-\cos\bar{\theta})^{2}}{N+N(N-1)\cos^{2}\bar{\theta}-[\nu^{2}-\nu(2N-1)]\sin^{2}\bar{\theta}}.
    \label{eq:remark3}
\end{align}
One can then show that squeezing is generated when the condition $\nu >(N-1)\cos^2(\bar{\theta}/2)$ is satisfied~\cite{SM}. In particular, a single detected photon ($\nu=1$) leads to postselected entanglement only for states close to the fully excited one, such that $\bar{\theta}>2\pi-2/\sqrt{N-1}$. 

\begin{figure}[b!]
\centering
\includegraphics[width=\columnwidth]{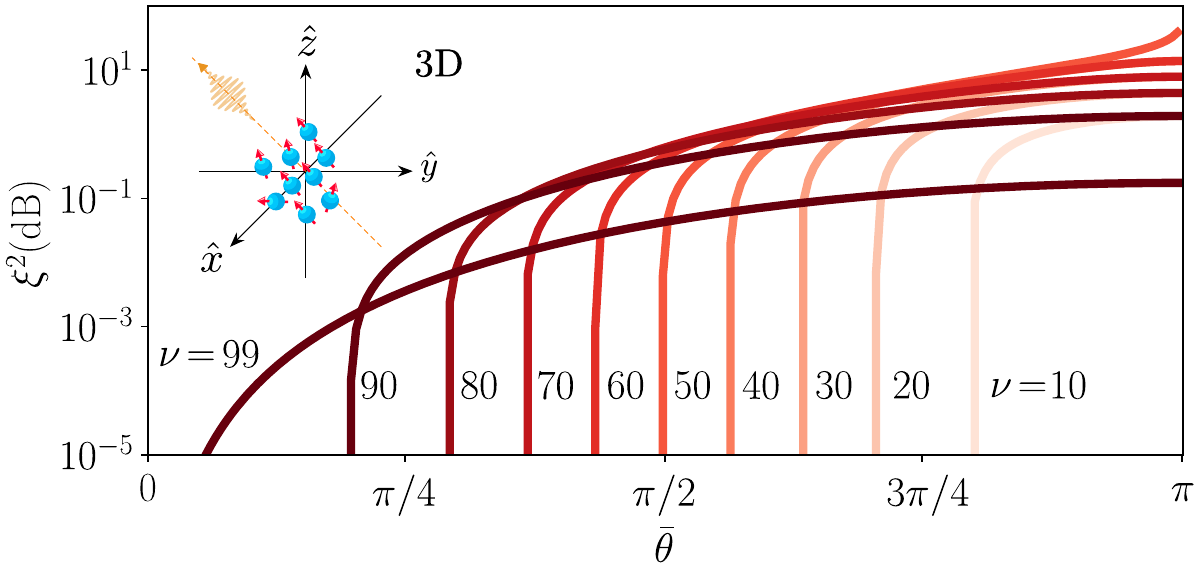}
\caption{ Squeezing parameter for a three-dimensional cloud of $N=100$ two-level emitters randomly distributed inside a sphere of radius $d=100(2\pi/k)$ initially prepared in state~\eqref{eq:no_coherence}. Entanglement is detected in all the region $0<\bar{\theta}\leq \pi$ if an enough number of photons are detected (here $\nu=10, 20,\hdots, 90, 99$, with optimal case corresponding to $\nu=N/2=50$). Note that as we get close to the ground state ($\bar{\theta}\xrightarrow{}0$), more photons are necessary to detect entanglement. }
\label{Fig:3}
\end{figure}    

The squeezing resulting from the detection of $\nu$ photons from a state of $N$ atoms with individual excited population $\sin^2(\bar{\theta}/2)$ is detailed in Fig.~\ref{Fig:3}. On the one hand, the strongest squeezing is reached in the limit $\bar{\theta}=\pi$ of a fully excited state, and for a photon number $\nu=N/2$, as in the case of CSSs. On the other hand, while initial states close to the fully excited one ($\bar{\theta}=\pi$) require few photons to generate spin squeezing, lower values of population require larger numbers $\nu$. Note the three-dimensional disordered nature of the system of emitters studied in this case, as compared to the one-dimensional chains investigated for CSSs and driven steady states, illustrates the universality of the postselected entanglement generation, in the sense that they do not depend on a particular geometry.

\begin{figure}[t!]
\centering
\includegraphics[width=\columnwidth]{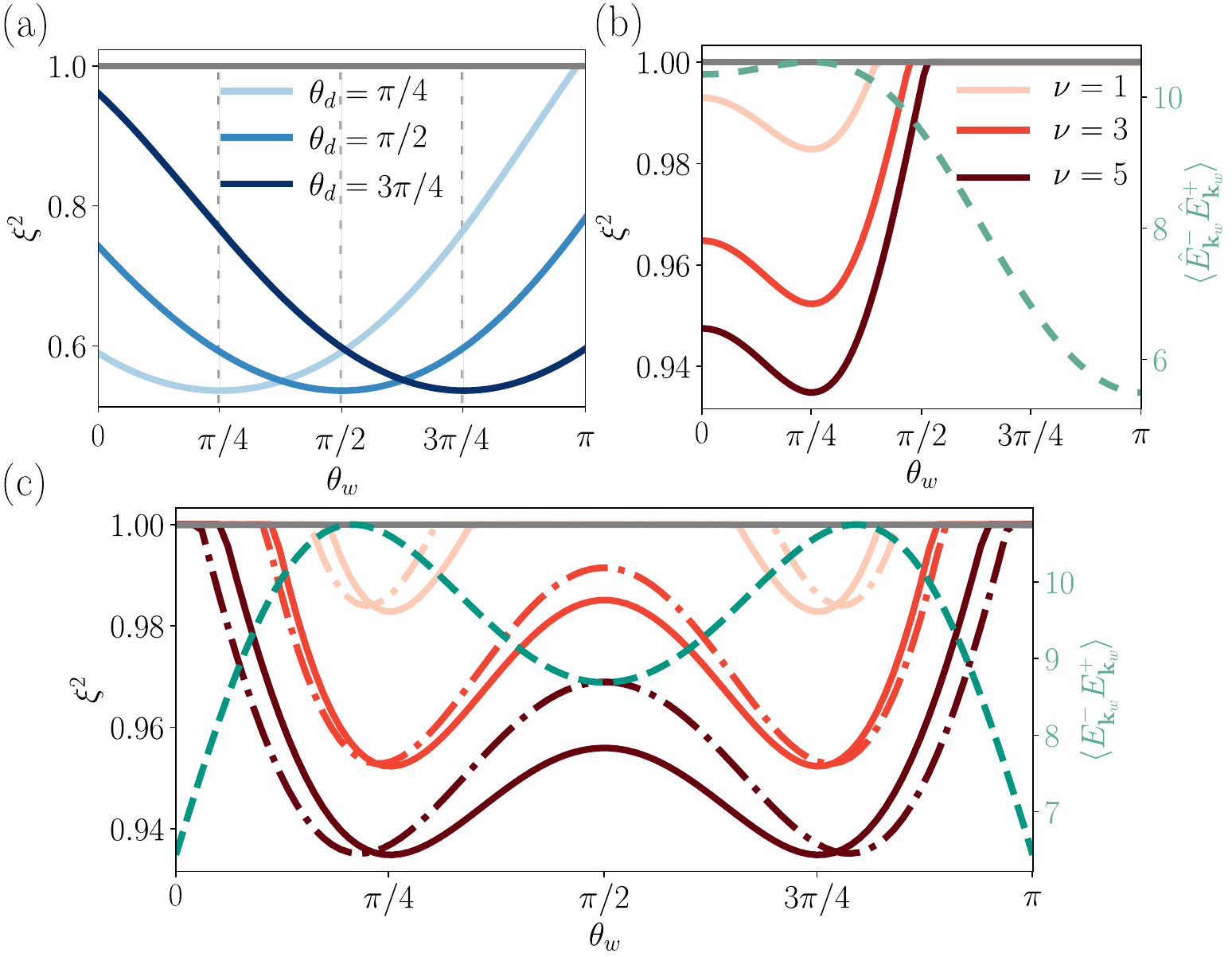}
\caption{Squeezing parameter for a 3D random configuration of $N=100$ two-level emitters inside a sphere of radius $d=100(2\pi/k)$: (a)  for the population state~\eqref{eq:no_coherence} with $\bar{\theta}=\pi/3$ and $\nu=50$ detected photons along three different directions $\theta_{d}$ (marked by the vertical dashed lines). The optimal direction $\theta_{w}$ to measure the squeezing is thus along the same direction of the photon detection, $\theta_{d}$. (b)-(c)  For $N=10$ atoms initially in CSS~\eqref{eq:pure} with $\theta=3\pi/4$ and $\mathbf{k}_{L}=\pi/4$. (b) 1D linear configuration along $z$-axis with $d=2\pi/k$ with all five photons detected along the drive direction $\mathbf{k}_{L}$.  (c)  2D circular configuration with radius $d=2\pi/k$, the photons are detected either in the same direction as the drive $\mathbf{k}_{L}$ (solid curve), or in 5 different angles (dash-dotted curves, with $\theta_1=0$, $\theta_2=\pi/3$, $\theta_3=\pi/2$, $\theta_4=3\pi/2$, $\theta_5=\pi$, always in the $xz$-plane). In (b-c), the dashed curve corresponds to the radiated intensity after a single photon detection (right axis).}
\label{Fig:4}
\end{figure}

{\it Spin squeezing orientation---}Until now, we have considered the same direction for the photon detection events and the squeezing measurement ($\mathbf{k}_d=\mathbf{k}_w$), as well as for initial states with optical coherences ($\mathbf{k}_d=\mathbf{k}_w=\mathbf{k}_L$): Let us now discuss why this is an ideal choice for the conditional generation of postselected spin squeezing. 

In Fig.~\ref{Fig:4}(a), for a 3D disordered system in an initial population state~\eqref{eq:no_coherence} we present the evolution of the squeezing parameter $\xi^2$ as a function of the angle $\theta_w$ of squeezing measurement, for three different directions $\theta_d$ of photon detection: While the initial state does not have any preferential direction of emission (the population states do not possess any optical coherence or phase terms), the strongest squeezing is systematically detected along the same direction as the photon detection, $\mathbf{k}_d=\mathbf{k}_w$ --- this feature could already be observed in Eq.~\eqref{xi_remark_2} for driven systems. This feature can be interpreted as the fact that these detection events create a collective spin along that specific direction $\theta_d$. In this  context, evaluating the squeezing along a different axis ($\mathbf{k}_d\neq\mathbf{k}_w$) is equivalent to characterizing the spin squeezing of a collective spin state along axes where the length spin $\langle \mathbf{J}_\parallel \rangle$ and its fluctuations $\langle \Delta \mathbf{J}_\perp \rangle$ are not maximum and minimum, respectively, thus not detecting the maximum spin squeezing of the state. 

Furthermore, when a phase is imprinted on the atomic cloud, such as for the coherent state~\eqref{eq:pure}, a preferential direction of emission arises, given by the drive. Then, it is precisely in that same direction that the strongest squeezing is created, by detecting the photons and measuring the squeezing ($\mathbf{k}_d=\mathbf{k}_w=\mathbf{k}_L$). This is illustrated in Fig.~\ref{Fig:4}(b), where the squeezing parameter $\xi^2$ is shown as a function of the angle of detection (considering $\mathbf{k}_d=\mathbf{k}_w$), for a two-dimensional ring of atoms: The strongest squeezing is reached for $\theta_w=\theta_L$. This shows that the coherent drive already imprints an optimal direction for the formation of the collective spin. Interestingly, this is also the emission of the maximum intensity, thanks to a constructive interference effect, see green dashed curve: It is thus the direction in which the photon(s) used to generate the conditional entanglement have the highest chance to be detected.

Finally, while detecting successive photons in the same direction may be challenging, let us now consider detecting them in different directions. In Fig.~\ref{Fig:4}(c), the squeezing obtained when considering different angles of detection is presented, for $\nu=1$, $3$ and $5$ detections (see dash-dotted curves): It exhibits only a small variation as compared to the same-direction case (plain curves). Note that we have checked that fully random angles (in the $4\pi\textrm{rad}^2$) provides similar results, with only the direction of the collective spin fluctuating. This highlights that the direction of detection of the photons is not critical to the formation of spin squeezing. It also suggests that other forms of detection, such as the detection in {\it any} direction (which is associated with the emission event for the atoms, and thus to $\hat{\sigma}^z$ operators rather than $\hat{\sigma}^-$), would also lead to the formation of squeezing.

{\it Conclusion---}In this work, we have shown how spin squeezing is generated by detecting photons from an initially separable state for large systems of quantum emitters. Accessed through far-field measurements of the electric field and its fluctuations, squeezing can be achieved even for initially mixed states, with successive detections acting as a purification process. The direction in which the photons are detected is optimal to characterize the squeezing, which shows that it is the direction in which the collective spin is created, with reduced fluctuations in the transverse directions. 

Note that this process bears some analogy with Dicke superradiance~\cite{Dicke_1954}, where the first emitted photon(s) imprints optical phases and thus a preferential direction of observation~\cite{Gross1982}. While our approach does not consider any dynamics, and thus no interaction between the emitters is involved, it would be interesting to investigate the role of cooperative emission, jointly with photon detection events, on the formation of squeezing. In particular, dipolar interactions have been shown to be able to induce scalable squeezing~\cite{Bornet_2023}, and may thus enhance the squeezing induced by photon detection.
 
\begin{acknowledgments}
{\it Acknowledgments---}The authors thank Lukáš Slodička and Robin Kaiser for fruitful discussions. They acknowledge the financial support of the São Paulo Research Foundation (FAPESP) (Grants No. 2024/05564-4, 2022/12382-4, 2018/15554-5, 2019/13143-0, 2019/12842-2, 2022/06449-9, 2023/07463-8, and 2023/03300-7), from the Brazilian CNPq (Conselho Nacional de Desenvolvimento Científico e Tecnológico), Grant No. 313632/2023-5. 
\end{acknowledgments}

\bibliography{bib.bib}
	

\onecolumngrid
\newpage
\begin{center}
{\large{ {\bf Supplemental Material for: \\ Postselected Entangled States by Photon Detection}}}

\setcounter{page}{1}

\vskip0.5\baselineskip{P. Rosario~\orcidlink{0000-0002-7628-7373},$^{1,2}$, A. Cidrim~\orcidlink{0000-0003-0007-2330},$^{1}$ and R. Bachelard~\orcidlink{0000-0002-6026-509X}$^{1}$}

\vskip0.5\baselineskip{ {\it $^{1}$Departamento de Física, Universidade Federal de São Carlos,\\ Rodovia Washington Luís, km 235 - SP-310, 13565-905 São Carlos, SP, Brazil}}
\vskip0.5\baselineskip{{\it $^{2}$CESQ/ISIS (UMR 7006), CNRS and Universit\'{e} de Strasbourg, 67000 Strasbourg, France}}

\end{center}

\appendix

\section{Single-photon Emission}

Before any quantum emission, the system is represented by a full separable quantum state without entanglement.
\begin{align}
     \hat{\rho}_{0} &=\bigotimes_{p=1}^{N}\hat{\rho}_{p}.
\end{align}
where each single emitter is explicitly given by
\begin{align}
    \hat{\rho}_{p}=\begin{pmatrix}   \rho_{p}^{ee} && \rho_{p}^{eg}  \\
    \rho_{p}^{ge} && \rho_{p}^{gg} 
\end{pmatrix}=\rho_{p}^{ee}\ketbra{e}{e}+\rho_{p}^{eg}\ketbra{e}{g}+\rho_{p}^{ge}\ketbra{g}{e}+\rho_{p}^{gg}\ketbra{g}{g},\ \ \text{with}\ \ \rho_{p}^{ee}+\rho_{p}^{gg}=1.
\end{align}
after a photon emission the quantum state takes the form
\begin{align}
    \hat{\rho}_1=\frac{\hat{E}^{+}_{\mathbf{k}_{d}} \hat{\rho}_{0} \hat{E}^{-}_{\mathbf{k}_{d}}}{\text{Tr}\left[\hat{E}^{+}_{\mathbf{k}_{d}} \hat{\rho}_{0} \hat{E}^{-}_{\mathbf{k}_{d}}\right]},
\end{align}
where the far-field single photon emission reads
\begin{align}
    \hat{E}^{+}_{\mathbf{k}_{d}}=\sum_{p=1}^{N}\me^{-\mi\mathbf{k}_{d}.\mathbf{r}_p}\hat{\sigma}^{-}_{p},
\end{align}
explicitly we get:
\begin{align}
     \hat{E}^{+}_{\mathbf{k}_{d}} \hat{\rho}_{0} \hat{E}^{-}_{\mathbf{k}_{d}} = \sum_{p=1}^{N}\bigotimes_{l\neq p}^{N}\hat{\rho}_{l}\otimes (\rho_{p}^{ee}\ketbra{g}{g})+\sum_{p=1}^{N}\sum_{q\neq p}^{N}\me^{-\mi\mathbf{k}_{d}.\mathbf{r}_p}\me^{\mi\mathbf{k}_{d}.\mathbf{r}_q}\bigotimes_{l\neq (p,q)}^{N}\hat{\rho}_{l}\otimes A_{p} \otimes B_{q},
     \label{eq:app_post_selected_1}
\end{align}
 with $A_{p}= \rho_{p}^{eg}\ketbra{g}{g}+\rho_{p}^{ee}\ketbra{g}{e}$,  $B_{q}=\rho_{q}^{ge}\ketbra{g}{g}+\rho_{q}^{ee}\ketbra{e}{g}$ and normalization factor 
\begin{align}
    F = \text{Tr}\left[\hat{E}^{+}_{\mathbf{k}_{d}} \hat{\rho}_{0} \hat{E}^{-}_{\mathbf{k}_{d}}\right]=\sum_{p=1}^{N}\rho_{p}^{ee}+\sum_{p=1}^{N}\sum_{q\neq p}^{N}\me^{-\mi\mathbf{k}_{d}.\mathbf{r}_p}\me^{\mi\mathbf{k}_{d}.\mathbf{r}_q}\rho_{p}^{eg}\rho_{q}^{ge},
\end{align}
\section{Electric Field Quadratures}
Considering the far-field limit, the photon emission quadratures are given by 
\begin{align}
&\hat{X}_{\mathbf{k}_{w}}=\hat{E}_{\mathbf{k}_{w}}^{+}+\hat{E}_{\mathbf{k}_{w}}^{-}=\sum_{s=1}^{N}\left(\me^{- \mi\mathbf{k}_{w}.\mathbf{r}_s}\hat{\sigma}^{-}_{s}+\me^{ \mi\mathbf{k}_{w}.\mathbf{r}_s}\hat{\sigma}^{+}_{s}\right)=\sum_{s=1}^{N}\hat{\sigma}_{\mathbf{k}_{w},s}^{x},\\
    &\hat{Y}_{\mathbf{k}_{w}}=\mi(\hat{E}_{\mathbf{k}_{w}}^{+}-\hat{E}_{\mathbf{k}_{w}}^{-})=\sum_{s=1}^{N}\mi\left( \me^{- \mi\mathbf{k}_{w}.\mathbf{r}_s}\hat{\sigma}^{-}_{s} - \me^{ \mi\mathbf{k}_{w}.\mathbf{r}_s}\hat{\sigma}^{+}_{s}\right)=\sum_{s=1}^{N}\hat{\sigma}_{\mathbf{k}_{w},s}^{y}.
\end{align}
and the inversion population operator
\begin{align}
    \hat{Z}=\sum_{s=1}^{N}\hat{\sigma}^{z}_{s},
\end{align}
we measure the quadratures along the same direction as the emitted photon, then $\mathbf{k}_{w}=\mathbf{k}_{d}$ and we get
\begin{align}
    &\nonumber  \langle \hat{X}_{\mathbf{k}_{d}} \rangle = \frac{1}{F}\sum_{s=1}^{N}\sum_{p\neq s}^{N}\rho_{p}^{ee}\left(\me^{-\mi\mathbf{k}_{d}.\mathbf{r}_s}\rho_{s}^{eg}+\me^{\mi\mathbf{k}_{d}.\mathbf{r}_s}\rho_{s}^{ge}\right)\\
    &+\frac{1}{F}\sum_{s=1}^{N}\sum_{q\neq s}^{N}\me^{\mi\mathbf{k}_{d}.\mathbf{r}_q}\rho_{s}^{ee} \rho_{q}^{ge}+\frac{1}{F}\sum_{s=1}^{N}\sum_{p\neq s}^{N}\me^{-\mi\mathbf{k}_{d}.\mathbf{r}_p}\rho_{s}^{ee} \rho_{p}^{eg}+ \frac{1}{F}\sum_{s=1}^{N}\sum_{p\neq s}^{N}\sum_{q\neq (s,p)}^{N}\me^{-\mi\mathbf{k}_{d}.\mathbf{r}_p}\me^{\mi\mathbf{k}_{d}.\mathbf{r}_q}\rho_{p}^{eg}\rho_{q}^{ge}(\me^{ -\mi\mathbf{k}_{d}.\mathbf{r}_s}\rho_{s}^{eg}+\me^{\mi\mathbf{k}_{d}.\mathbf{r}_s}\rho_{s}^{ge}),\\
    &\nonumber  \langle \hat{Y}_{\mathbf{k}_{d}} \rangle = \frac{\mi}{F}\sum_{s=1}^{N}\sum_{p\neq s}^{N}\rho_{p}^{ee}\left(\me^{-\mi\mathbf{k}_{d}.\mathbf{r}_s}\rho_{s}^{eg}-\me^{\mi\mathbf{k}_{d}.\mathbf{r}_s}\rho_{s}^{ge}\right)\\
    &-\frac{\mi}{F}\sum_{s=1}^{N}\sum_{q\neq s}^{N}\me^{\mi\mathbf{k}_{d}.\mathbf{r}_q}\rho_{s}^{ee} \rho_{q}^{ge}+\frac{\mi}{F}\sum_{s=1}^{N}\sum_{p\neq s}^{N}\me^{-\mi\mathbf{k}_{d}.\mathbf{r}_p}\rho_{s}^{ee} \rho_{p}^{eg}+ \frac{\mi}{F}\sum_{s=1}^{N}\sum_{p\neq s}^{N}\sum_{q\neq (s,p)}^{N}\me^{-\mi\mathbf{k}_{d}.\mathbf{r}_p}\me^{\mi\mathbf{k}_{d}.\mathbf{r}_q}\rho_{p}^{eg}\rho_{q}^{ge}(\me^{ -\mi\mathbf{k}_{d}.\mathbf{r}_s}\rho_{s}^{eg}-\me^{\mi\mathbf{k}_{d}.\mathbf{r}_s}\rho_{s}^{ge}),\\
    &\nonumber  \langle \hat{Z} \rangle = -\frac{1}{F}\sum_{s=1}^{N}\rho_{s}^{ee}+\frac{1}{F}\sum_{s=1}^{N}\sum_{p\neq s}^{N}\rho_{p}^{ee}(\rho_{s}^{ee}-\rho_{s}^{gg})\\
    &-\frac{1}{F}\sum_{s=1}^{N}\sum_{q\neq s}^{N}\me^{-\mi\mathbf{k}_{d}.\mathbf{r}_s}\me^{\mi\mathbf{k}_{d}.\mathbf{r}_q}\rho_{s}^{eg}\rho_{q}^{ge}-\frac{1}{F}\sum_{s=1}^{N}\sum_{p\neq s}^{N}\me^{-\mi\mathbf{k}_{d}.\mathbf{r}_p}\me^{\mi\mathbf{k}_{d}.\mathbf{r}_s}\rho_{s}^{ge}\rho_{p}^{eg}+\frac{1}{F}\sum_{s=1}^{N}\sum_{p\neq s}^{N}\sum_{q\neq (s,p)}^{N}\me^{-\mi\mathbf{k}_{d}.\mathbf{r}_p}\me^{\mi\mathbf{k}_{d}.\mathbf{r}_q}\rho_{p}^{eg}\rho_{q}^{ge}(\rho_{s}^{ee}-\rho_{s}^{gg}).
\end{align}
similarly
\begin{align}
     \nonumber \langle \hat{X}^{2}_{\mathbf{k}_{d}} \rangle &=N + \frac{1}{F}\sum_{s=1}^{N}\sum_{l\neq s}^{N}\sum_{p\neq (s,l)}^{N}\rho_{p}^{ee}(\me^{-\mi\mathbf{k}_{d}.\mathbf{r}_s}\rho_{s}^{eg}+\me^{\mi\mathbf{k}_{d}.\mathbf{r}_s}\rho_{s}^{ge})(\me^{-\mi\mathbf{k}_{d}.\mathbf{r}_l}\rho_{l}^{eg}+\me^{\mi\mathbf{k}_{d}.\mathbf{r}_l}\rho_{l}^{ge}) + \frac{2}{F}\sum_{s=1}^{N}\sum_{l\neq s}^{N}\rho_{s}^{ee} \rho_{l}^{ee} \\
    &\nonumber + \frac{2}{F}\sum_{s=1}^{N}\sum_{l\neq s}^{N}\sum_{q\neq (s,l)}^{N}\me^{\mi\mathbf{k}_{d}.\mathbf{r}_q}\rho_{s}^{ee} (\me^{-\mi\mathbf{k}_{d}.\mathbf{r}_l}\rho_{l}^{eg}+\me^{ \mi\mathbf{k}_{d}.\mathbf{r}_l}\rho_{l}^{ge})\rho_{q}^{ge} + \frac{2}{F}\sum_{s=1}^{N}\sum_{l\neq s}^{N}\sum_{p\neq (s,l)}^{N} \me^{-\mi\mathbf{k}_{d}.\mathbf{r}_p}\rho_{s}^{ee}(\me^{-\mi\mathbf{k}_{d}.\mathbf{r}_l}\rho_{l}^{eg}+\me^{ \mi\mathbf{k}_{d}.\mathbf{r}_l}\rho_{l}^{ge})\rho_{p}^{eg}\\
    & + \frac{1}{F}\sum_{s=1}^{N}\sum_{l\neq s}^{N}\sum_{p\neq (s,l)}^{N}\sum_{q\neq (p,s,l)}^{N}\me^{-\mi\mathbf{k}_{d}.\mathbf{r}_p}\me^{\mi\mathbf{k}_{d}.\mathbf{r}_q}(\me^{-\mi\mathbf{k}_{d}.\mathbf{r}_s}\rho_{s}^{eg}+\me^{\mi\mathbf{k}_{d}.\mathbf{r}_s}\rho_{s}^{ge})(\me^{-\mi\mathbf{k}_{d}.\mathbf{r}_l}\rho_{l}^{eg}+\me^{\mi\mathbf{k}_{d}.\mathbf{r}_l}\rho_{l}^{ge})\rho_{p}^{eg}\rho_{q}^{ge},
\end{align}
\begin{align}
     \nonumber \langle \hat{Y}^{2}_{\mathbf{k}_{d}} \rangle &= N  -\frac{1}{F}\sum_{s=1}^{N}\sum_{l\neq s}^{N}\sum_{p\neq (s,l)}^{N}\rho_{p}^{ee}(\me^{-\mi\mathbf{k}_{d}.\mathbf{r}_s}\rho_{s}^{eg}-\me^{\mi\mathbf{k}_{d}.\mathbf{r}_s}\rho_{s}^{ge})(\me^{-\mi\mathbf{k}_{d}.\mathbf{r}_l}\rho_{l}^{eg}-\me^{\mi\mathbf{k}_{d}.\mathbf{r}_l}\rho_{l}^{ge}) + \frac{2}{F}\sum_{s=1}^{N}\sum_{l\neq s}^{N}\rho_{s}^{ee} \rho_{l}^{ee} \\
    &\nonumber + \frac{2}{F}\sum_{s=1}^{N}\sum_{l\neq s}^{N}\sum_{q\neq (s,l)}^{N}\me^{\mi\mathbf{k}_{d}.\mathbf{r}_q}\rho_{s}^{ee} (\me^{-\mi\mathbf{k}_{d}.\mathbf{r}_l}\rho_{l}^{eg}-\me^{ \mi\mathbf{k}_{d}.\mathbf{r}_l}\rho_{l}^{ge})\rho_{q}^{ge} - \frac{2}{F}\sum_{s=1}^{N}\sum_{l\neq s}^{N}\sum_{p\neq (s,l)}^{N} \me^{-\mi\mathbf{k}_{d}.\mathbf{r}_p}\rho_{s}^{ee} (\me^{-\mi\mathbf{k}_{d}.\mathbf{r}_l}\rho_{l}^{eg}-\me^{ \mi\mathbf{k}_{d}.\mathbf{r}_l}\rho_{l}^{ge})\rho_{p}^{eg}\\
    & - \frac{1}{F}\sum_{s=1}^{N}\sum_{l\neq s}^{N}\sum_{p\neq (s,l)}^{N}\sum_{q\neq (p,s,l)}^{N}\me^{-\mi\mathbf{k}_{d}.\mathbf{r}_p}\me^{\mi\mathbf{k}_{d}.\mathbf{r}_q}(\me^{-\mi\mathbf{k}_{d}.\mathbf{r}_s}\rho_{s}^{eg}-\me^{ \mi\mathbf{k}.\mathbf{r}_s}\rho_{s}^{ge})(\me^{-\mi\mathbf{k}_{d}.\mathbf{r}_l}\rho_{l}^{eg}-\me^{ \mi\mathbf{k}_{d}.\mathbf{r}_l}\rho_{l}^{ge})\rho_{p}^{eg}\rho_{q}^{ge},
\end{align}
\begin{align}
    \nonumber \langle \hat{Z}^{2}\rangle &=N -\frac{1}{F}\sum_{s=1}^{N}\sum_{l\neq s}^{N}\rho_{s}^{ee}(\rho_{l}^{ee}-\rho_{l}^{gg})-\frac{1}{F}\sum_{s=1}^{N}\sum_{l\neq s}^{N}\rho_{l}^{ee}(\rho_{s}^{ee}-\rho_{s}^{gg})+ \frac{1}{F}\sum_{s=1}^{N}\sum_{l\neq s}^{N}\sum_{p\neq (s,l)}^{N}(\rho_{s}^{ee}-\rho_{s}^{gg})(\rho_{l}^{ee}-\rho_{l}^{gg})\rho_{p}^{ee}\\
    &\nonumber + \frac{2}{F}\sum_{s=1}^{N}\sum_{l\neq s}^{N}\me^{-\mi\mathbf{k}_{d}.\mathbf{r}_s}\me^{\mi\mathbf{k}_{d}.\mathbf{r}_l}\rho_{s}^{eg}\rho_{l}^{ge} - \frac{4}{F}\sum_{s=1}^{N}\sum_{l\neq s}^{N}\sum_{q\neq (s,l)}^{N}\me^{-\mi\mathbf{k}_{d}.\mathbf{r}_s}\me^{\mi\mathbf{k}_{d}.\mathbf{r}_q}\rho_{s}^{eg}(\rho_{l}^{ee}-\rho_{l}^{gg})\rho_{q}^{ge}\\
    & + \frac{1}{F}\sum_{s=1}^{N}\sum_{l\neq s}^{N}\sum_{p\neq (s,l)}^{N}\sum_{q\neq (p,s,l)}^{N}\me^{-\mi\mathbf{k}_{d}.\mathbf{r}_p}\me^{\mi\mathbf{k}_{d}.\mathbf{r}_q}(\rho_{s}^{ee}-\rho_{s}^{gg})(\rho_{l}^{ee}-\rho_{l}^{gg})\rho_{p}^{eg}\rho_{q}^{ge}.
\end{align}
\section{Drive direction improves entanglement detection}
In this section, we show that considering a quantum state with relative phase induced by a laser drive, generates optical phase independent expected values (and then spin squeezing parameter) after a single photon detection. It implies that measuring the photon along the drive direction corresponds to the best case scenario when measuring the electric field.

Let us consider the family of quantum states
\begin{align}
    \ket{\psi}=\bigotimes_{p=1}^{N}\left[\cos\left(\frac{\theta}{2}\right)\ket{g}+\sin\left(\frac{\theta}{2}\right)e^{i\mathbf{k}_{L}.\mathbf{r}_{p}}\ket{e}\right].
\end{align}
the expected values will take the form:
\begin{align}
     \nonumber &\langle \hat{X}_{\mathbf{k}} \rangle=\frac{1}{F}\rho_{ee}\left(\rho_{eg}+\rho_{ge}\right)N(N-1)+\frac{1}{F}\rho_{ee} \rho_{ge}N(N-1)+\frac{1}{F}\rho_{ee} \rho_{eg}N(N-1)+\frac{1}{F}\rho_{eg}\rho_{ge}(\rho_{eg}+\rho_{ge})N(N-1)(N-2),\\
    \nonumber &\langle \hat{Y}_{\mathbf{k}} \rangle=0,\\
     \nonumber &\langle \hat{Z} \rangle =-\frac{1}{F}N\rho_{ee}+\frac{1}{F}N(N-1)\rho_{ee}(\rho_{ee}-\rho_{gg})-\frac{1}{F}\rho_{eg}\rho_{ge}N(N-1)-\frac{1}{F}\rho_{ge}\rho_{eg}N(N-1)\\
     &\nonumber+\frac{1}{F}\rho_{eg}\rho_{ge}(\rho_{ee}-\rho_{gg})N(N-1)(N-2).
\end{align}
non linear expected values

\begin{align}
    &\nonumber \langle \hat{X}^{2}_{\mathbf{k}} \rangle =N+\frac{1}{F}N(N-1)(N-2)\rho_{ee}(\rho_{eg}+\rho_{ge})(\rho_{eg}+\rho_{ge}) + \frac{2}{F}N(N-1)\rho_{ee} \rho_{ee}+\frac{2}{F}N(N-1)(N-2)\rho_{ee} (\rho_{eg}+\rho_{ge})\rho_{ge}\\
    &\nonumber+\frac{2}{F}N(N-1)(N-2)\rho_{ee}(\rho_{eg}+\rho_{ge})\rho_{eg}+\frac{1}{F}N(N-1)(N-2)(N-3)(\rho_{eg}+\rho_{ge})(\rho_{eg}+\rho_{ge})\rho_{eg}\rho_{ge},\\
    &\nonumber \langle \hat{Y}^{2}_{\mathbf{k}} \rangle =N-\frac{1}{F}N(N-1)(N-2)\rho_{ee}(\rho_{eg}-\rho_{ge})(\rho_{eg}-\rho_{ge}) + \frac{2}{F}N(N-1)\rho_{ee} \rho_{ee}+\frac{2}{F}N(N-1)(N-2)\rho_{ee} (\rho_{eg}-\rho_{ge})\rho_{ge}\\
    &\nonumber-\frac{2}{F}N(N-1)(N-2)\rho_{ee}(\rho_{eg}-\rho_{ge})\rho_{eg}-\frac{1}{F}N(N-1)(N-2)(N-3)(\rho_{eg}-\rho_{ge})(\rho_{eg}-\rho_{ge})\rho_{eg}\rho_{ge},\\
    &\nonumber \langle \hat{Z}^{2}\rangle =N-\frac{1}{F}N(N-1)\rho_{ee}(\rho_{ee}-\rho_{gg})-\frac{1}{F}N(N-1)\rho_{ee}(\rho_{ee}-\rho_{gg})+\frac{1}{F}N(N-1)(N-2)(\rho_{ee}-\rho_{gg})(\rho_{ee}-\rho_{gg})\rho_{ee}\\
    &\nonumber +\frac{2}{F}N(N-1)\rho_{eg}\rho_{ge}-\frac{4}{F}N(N-1)(N-2)\rho_{eg}(\rho_{ee}-\rho_{gg})\rho_{ge}+\frac{1}{F}N(N-1)(N-2)(N-3)(\rho_{ee}-\rho_{gg})(\rho_{ee}-\rho_{gg})\rho_{eg}\rho_{ge}.
\end{align}

where $\rho_{ee}=\sin^{2}(\theta/2)$, $\rho_{gg}=\cos^{2}(\theta/2)$, $\rho_{eg}=\rho_{ge}=\sin(\theta/2)\cos(\theta/2)$ and not optical phases are present.
\section{Postselected State Intensity}
The intensity of the scattered electric field can be obtained as
\begin{align}
    I=\langle \hat{E}_{\mathbf{k}_{w}}^{-}\hat{E}_{\mathbf{k}_{w}}^{+}\rangle.
\end{align}
for the postselected quantum state in Eq.\eqref{eq:app_post_selected_1} it reads explicitly
\begin{align}
    \nonumber  I &= \frac{1}{F}\sum_{s=1}^{N}\sum_{n\neq s}^{N}\rho_{s}^{ee}\rho_{n}^{ee}+ \frac{1}{F}\sum_{p=1}^{N}\sum_{q\neq p}^{N}\sum_{n\neq (p,q)}^{N}e^{-i\mathbf{k}_{d}.\mathbf{r}_p}e^{i\mathbf{k}_{d}.\mathbf{r}_q}\rho_{n}^{ee}\rho_{p}^{eg}\rho_{q}^{ge}+ \frac{1}{F}\sum_{p=1}^{N}\sum_{n\neq p}^{N}\sum_{n'\neq (p,n)}^{N}e^{i\mathbf{k}_{w}.\mathbf{r}_n}e^{-i\mathbf{k}_{w}.\mathbf{r}_{n'}}\rho_{p}^{ee}\rho_{n'}^{eg}\rho_{n}^{ge}\\
    &\nonumber+\frac{1}{F}\sum_{p=1}^{N}\sum_{q\neq q}^{N}e^{-i\mathbf{k}_{d}.\mathbf{r}_p}e^{i\mathbf{k}_{d}.\mathbf{r}_q}e^{-i\mathbf{k}_{w}.\mathbf{r}_q}e^{i\mathbf{k}_{w}.\mathbf{r}_p}\rho_{p}^{ee}\rho_{q}^{ee}+\frac{1}{F}\sum_{p=1}^{N}\sum_{q\neq q}^{N}\sum_{n\neq (p,q)}^{N}e^{-i\mathbf{k}_{d}.\mathbf{r}_p}e^{i\mathbf{k}_{d}.\mathbf{r}_q}e^{-i\mathbf{k}_{w}.\mathbf{r}_q}e^{i\mathbf{k}_{w}.\mathbf{r}_n}\rho_{n}^{ge}\rho_{p}^{eg}\rho_{q}^{ee}\\
    &\nonumber+\frac{1}{F}\sum_{p=1}^{N}\sum_{q\neq q}^{N}\sum_{n'\neq (p,q)}^{N}e^{-i\mathbf{k}_{d}.\mathbf{r}_p}e^{i\mathbf{k}_{d}.\mathbf{r}_q}e^{i\mathbf{k}_{w}.\mathbf{r}_p}e^{-i\mathbf{k}_{w}.\mathbf{r}_{n'}}\rho_{n'}^{eg}\rho_{q}^{ge}\rho_{p}^{ee}\\
    &+\frac{1}{F}\sum_{p=1}^{N}\sum_{q\neq p}^{N}\sum_{n\neq (p,q)}^{N}\sum_{n'\neq (n,p,q)}^{N}e^{-i\mathbf{k}_{d}.\mathbf{r}_p}e^{i\mathbf{k}_{d}.\mathbf{r}_q}e^{-i\mathbf{k}_{w}.\mathbf{r}_n}e^{i\mathbf{k}_{w}.\mathbf{r}_{n'}}\rho_{p}^{eg}\rho_{q}^{ge}\rho_{n}^{eg}\rho_{n'}^{ge}
\end{align}
note that for the initial quantum state in Eq.\eqref{eq:driven_state}, if the quadratures are measured along the detected photon direction $\mathbf{k}_{w}=\mathbf{k}_{d}$ and the detected photons are along the drive direction $\mathbf{k}_{L}$, the intensity is maximum, since all the optical phases disappear.
\section{Zero-coherence (Population) Quantum States}
Let us consider the initial quantum state, where no quantum coherences are initially present
\begin{align}
    \hat{\rho}_{0}=\frac{1}{2^{N}}\bigotimes_{p=1}^{N}\left(\1_{p}+v^{z}_{p}\hat{\sigma}^{z}_{p}\right),
\end{align}
a smart  way of writing $\hat{\rho}_{0}$ is given by
\begin{align}
    \nonumber \hat{\rho}_{0}&=\frac{1}{2^{N}}\bigotimes_{p=1}^{N}\1_{p}+\frac{1}{2^{N}}\sum_{\beta=1}^{N}\frac{1}{\beta!}\left(\sum_{p_{1}=1}^{N}\sum_{p_{2}\neq p_{1}}^{N}\hdots\sum_{p_{\beta}\neq (p_{\beta-1},\hdots,p_{1})}^{N}\left(\prod_{j=1}^{\beta}v^{z}_{p_{j}}\hat{\sigma}^{z}_{p_{j}}\right)\right),
\end{align}
 So, the quantum state after $n$-photon emissions is given by
\begin{align}
    \hat{\rho}_{n}=\frac{\left(\hat{E}_{\mathbf{k}_{d}}^{+}\right)^{n}\hat{\rho}_{0}\left(\hat{E}^{-}_{\mathbf{k}_{d}}\right)^{n}}{\text{Tr}\left[\left(\hat{E}_{\mathbf{k}_{d}}^{+}\right)^{n}\hat{\rho}_{0}\left(\hat{E}_{\mathbf{k}_{d}}^{-}\right)^{n}\right]},
\end{align}
we focus on the case where all the photons are emitted along the same direction $\mathbf{k}_{d}$ but the electric field can be measured in a different one $\mathbf{k}_{w}$. The number of photons emission is given by $1\leq n\leq N-1$ and we notice that the last expression can be written as:
\begin{align}
     \hat{\rho}_{n}\propto \left(\sum_{a_{1}=1}^{N}\sum_{a_{2}\neq a_{1}}^{N}\hdots \sum_{a_{n}\neq (a_{n-1},\hdots,a_{1})}^{N}\hat{\sigma}^{-}_{a_{1}}\hat{\sigma}^{-}_{a_{2}}\hdots \hat{\sigma}^{-}_{a_{n}}\right)\hat{\rho}_{0}\left(\sum_{b_{1}=1}^{N}\sum_{b_{2}\neq b_{1}}^{N}\hdots \sum_{b_{n}\neq (b_{n-1},\hdots,b_{1})}^{N}\hat{\sigma}^{+}_{b_{1}}\hat{\sigma}^{+}_{b_{2}}\hdots \hat{\sigma}^{+}_{b_{n}}\right),
\end{align}
after some algebra, the normalization factor reads
\begin{align}
    \text{Tr}\left[\left(\hat{E}_{\mathbf{k}_{d}}^{+}\right)^{n}\hat{\rho}_{0}\left(\hat{E}_{\mathbf{k}_{d}}^{-}\right)^{n}\right]=n!2^{N-n}\left(\prod_{k=0}^{n-1}(N-k)+\sum_{\alpha=1}^{n}\begin{pmatrix}
 n\\
\alpha  
\end{pmatrix}\left[\sum_{b_{1}=1}^{N}\sum_{b_{2}\neq b_{1}}^{N}\hdots \sum_{b_{n}\neq (b_{n-1},\hdots,b_{1})}^{N}\left(\prod_{j=1}^{\alpha}v^{z}_{b_{j}}\right)\right]\right),
\end{align}
It is possible to show that
\begin{align}
    \langle \hat{X}_{\mathbf{k}_{w}}\rangle=\langle \hat{Y}_{\mathbf{k}_{w}}\rangle=0,
\end{align}
Similarly (without normalization)
\begin{align}
    \nonumber \langle \hat{Z}\rangle &\propto-n2^{N-n}n!\left(\prod_{k=0}^{n-1}(N-k)+\sum_{\alpha=1}^{n}\begin{pmatrix}
 n\\
\alpha  
\end{pmatrix}\left[\sum_{b_{1}=1}^{N}\sum_{b_{2}\neq b_{1}}^{N}\hdots \sum_{b_{n}\neq (b_{n-1},\hdots,b_{1})}^{N}\left(\prod_{j=1}^{\alpha}v^{z}_{b_{j}}\right)\right]\right)\\
&\nonumber+n!2^{N-n}\sum_{b_{1}=1}^{N}\sum_{b_{2}\neq b_{1}}^{N}\hdots \sum_{b_{n}\neq (b_{n-1},\hdots,b_{1})}^{N}\sum_{p\neq (b_{n},b_{n-1},\hdots,b_{1})}^{N}v^{z}_{p}\\
&+n!2^{N-n}\sum_{\alpha=1}^{n}\begin{pmatrix}
 n\\
\alpha  
\end{pmatrix}\left(\sum_{b_{1}=1}^{N}\sum_{b_{2}\neq b_{1}}^{N}\hdots \sum_{b_{n}\neq (b_{n-1},\hdots,b_{1})}^{N}\sum_{p\neq (b_{n},\hdots,b_{1})}^{N}\left(\prod_{j=1}^{\alpha}v^{z}_{b_{j}}\right)v^{z}_{p}\right),
\end{align}
\begin{align}
     \nonumber\langle (\hat{X}_{\mathbf{k}_{w}})^{2}\rangle &\propto Nn!2^{N-n}\prod_{g=0}^{n-1}(N-g)+n!2^{N-n}N\sum_{\alpha=1}^{n}\begin{pmatrix}
 n\\
\alpha  
\end{pmatrix}\left[\sum_{b_{1}=1}^{N}\sum_{b_{2}\neq b_{1}}^{N}\hdots \sum_{b_{n}\neq (b_{n-1},\hdots,b_{1})}^{N}\left(\prod_{j=1}^{\alpha}v^{z}_{b_{j}}\right)\right]\\
     &\nonumber +n!2^{N-n}n \left(\delta_{1n}+(1-\delta_{1n})\prod_{g=2}^{n}(N-g)\right)\sum_{b_{1}=1}^{N}\sum_{b_{2}\neq b_{1}}^{N}\me^{-\mi(\mathbf{k}_{d}-\mathbf{k}_{w}).\mathbf{r}_{b_{1}}}\me^{\mi(\mathbf{k}_{d}-\mathbf{k}_{w}).\mathbf{r}_{b_{2}}}\\
     &\nonumber+n!2^{N-n}n \sum_{\alpha=1}^{n}\begin{pmatrix}
 n\\
\alpha  
\end{pmatrix}\left[\sum_{b_{1}=1}^{N}\sum_{b_{2}\neq b_{1}}^{N}\hdots \sum_{b_{n}\neq (b_{n-1},\hdots,b_{1})}^{N}\sum_{b_{n+1}\neq (b_{n},\hdots,b_{1})}^{N}\me^{-\mi(\mathbf{k}_{d}-\mathbf{k}_{w}).\mathbf{r}_{b_{1}}}\me^{\mi(\mathbf{k}_{d}-\mathbf{k}_{w}).\mathbf{r}_{b_{2}}}\left(\prod_{j=1}^{\alpha}v^{z}_{b_{j}}\right)\right]\\
&\nonumber+n!2^{N-n}n\sum_{b_{1}=1}^{N}\sum_{b_{2}\neq b_{1}}^{N}\hdots \sum_{b_{n}\neq (b_{n-1},\hdots,b_{1})}^{N}\sum_{b_{n+1}\neq (b_{n},b_{n-1},\hdots,b_{1})}^{N}\me^{-\mi(\mathbf{k}_{d}-\mathbf{k}_{w}).\mathbf{r}_{b_{n+1}}}\me^{\mi(\mathbf{k}_{d}-\mathbf{k}_{w}).\mathbf{r}_{b_{n}}}v^{z}_{b_{n+1}}\\
&+n!2^{N-n}n\sum_{\alpha=1}^{n}\begin{pmatrix}
 n\\
\alpha  
\end{pmatrix}\left(\sum_{b_{1}=1}^{N}\sum_{b_{2}\neq b_{1}}^{N}\hdots \sum_{b_{n}\neq (b_{n-1},\hdots,b_{1})}^{N}\sum_{b_{n+1}\neq (b_{n},\hdots,b_{1})}^{N}\me^{-\mi(\mathbf{k}_{d}-\mathbf{k}_{w}).\mathbf{r}_{b_{n+1}}}\me^{\mi(\mathbf{k}_{d}-\mathbf{k}_{w}).\mathbf{r}_{b_{n}}}\left(\prod_{j=1}^{\alpha}v^{z}_{b_{j}}\right)v^{z}_{b_{n+1}}\right),
\end{align}
It is possible to notice that $\langle (\hat{X}_{\mathbf{k}_{w}})^{2}\rangle=\langle (\hat{Y}_{\mathbf{k}_{w}})^{2}\rangle$ and
\begin{align}
    \nonumber &\langle \hat{Z}^{2}\rangle \propto Nn!2^{N-n}\prod_{k=0}^{n-1}(N-k)+Nn!2^{N-n}\sum_{\alpha=1}^{n}\begin{pmatrix}
 n\\
\alpha  
\end{pmatrix}\left[\sum_{b_{1}=1}^{N}\sum_{b_{2}\neq b_{1}}^{N}\hdots \sum_{b_{n}\neq (b_{n-1},\hdots,b_{1})}^{N}\left(\prod_{j=1}^{\alpha}v^{z}_{b_{j}}\right)\right]\\
    &\nonumber+n(n-1)n!2^{N-n}\prod_{k=0}^{n-1}(N-k)+n(n-1)n!2^{N-n}\sum_{\alpha=1}^{n}\begin{pmatrix}
 n\\
\alpha  
\end{pmatrix}\left[\sum_{b_{1}=1}^{N}\sum_{b_{2}\neq b_{1}}^{N}\hdots \sum_{b_{n}\neq (b_{n-1},\hdots,b_{1})}^{N}\left(\prod_{j=1}^{\alpha}v^{z}_{b_{j}}\right)\right]\\
&\nonumber -2n!2^{N-n}n\sum_{\alpha=1}^{n}\begin{pmatrix}
 n\\
\alpha  
\end{pmatrix}\left(\sum_{b_{1}=1}^{N}\sum_{b_{2}\neq b_{1}}^{N}\hdots \sum_{b_{n}\neq (b_{n-1},\hdots,b_{1})}^{N}\sum_{p\neq (b_{n},\hdots,b_{1})}^{N}\left(\prod_{j=1}^{\alpha}v^{z}_{b_{j}}\right)v^{z}_{p}\right)\\
&\nonumber-2n!2^{N-n}n\sum_{b_{1}=1}^{N}\sum_{b_{2}\neq b_{1}}^{N}\hdots \sum_{b_{n}\neq (b_{n-1},\hdots,b_{1})}^{N}\sum_{p\neq (b_{n},b_{n-1},\hdots,b_{1})}^{N}v^{z}_{p}\\
&\nonumber+n!2^{N-n} \left(1-\delta_{2,N}\right)\left(1-\delta_{n,N-1}\right)\sum_{b_{1}=1}^{N}\sum_{b_{2}\neq b_{1}}^{N}\hdots \sum_{b_{n}\neq (b_{n-1},\hdots,b_{1})}^{N}\sum_{p_{1}\neq (b_{n},b_{n-1},\hdots,b_{1})}^{N}\sum_{p_{2}\neq (p_{1},b_{n},\hdots,b_{1})}^{N}v^{z}_{p_{1}}v^{z}_{p_{2}}\\
&+n!2^{N-n}\left(1-\delta_{2,N}\right)\left(1-\delta_{n,N-1}\right)\sum_{\alpha=1}^{n}\begin{pmatrix}
 n\\
\alpha  
\end{pmatrix}\left(\sum_{b_{1}=1}^{N}\sum_{b_{2}\neq b_{1}}^{N}\hdots \sum_{b_{n}\neq (b_{n-1},\hdots,b_{1})}^{N}\sum_{p_{1}\neq (b_{n},b_{n-1},\hdots,b_{1})}^{N}\sum_{p_{2}\neq (p_{1},b_{n},\hdots,b_{1})}^{N}\left(\prod_{j=1}^{\alpha} v^{z}_{b_{j}}\right)v^{z}_{p_{1}}v^{z}_{p_{2}}\right),
\end{align}
for the case $v^{z}_{1}=v^{z}_{2}=\hdots=v^{z}_{N}=v^{z}$, it reduces to
\begin{align}
&\langle \hat{Z}\rangle=-n+v^{z}(N-n),\\
&\nonumber \langle (\hat{X}_{\mathbf{k}_{w}})^{2}\rangle =N+\frac{1}{\prod_{g=0}^{n-1}(N-g)}n\left(\delta_{1n}+(1-\delta_{1n})\prod_{g=2}^{n}(N-g)\right)\sum_{b_{1}=1}^{N}\sum_{b_{2}\neq b_{1}}^{N}\me^{-\mi(\mathbf{k}_{d}-\mathbf{k}_{w}).\mathbf{r}_{b_{1}}}\me^{\mi(\mathbf{k}_{d}-\mathbf{k}_{w}).\mathbf{r}_{b_{2}}}\\
     &+\frac{1}{\prod_{g=0}^{n-1}(N-g)}n v^{z}\left(\delta_{1n}+(1-\delta_{1n})\prod_{g=2}^{n}(N-g)\right)\sum_{b_{1}=1}^{N}\sum_{b_{2}\neq b_{1}}^{N}\me^{-\mi(\mathbf{k}_{d}-\mathbf{k}_{w}).\mathbf{r}_{b_{1}}}\me^{\mi(\mathbf{k}_{d}-\mathbf{k}_{w}).\mathbf{r}_{b_{2}}},\\
    &\langle \hat{Z}^{2}\rangle= N+n(n-1)-2n(N-n)v^{z}+\left(1-\delta_{2,N}\right)\left(1-\delta_{n,N-1}\right)(N-n)(N-(n+1))(v^{z})^{2}.
\end{align}
It is equivalent to
\begin{align}
    &\langle \hat{Z}\rangle=-n+v_{z}(N-n),\\
    &\nonumber \langle (\hat{X}_{\mathbf{k}_{w}})^{2}\rangle=N+(1+v^{z})\frac{n(N-n)}{N(N-1)}\sum_{b_{1}=1}^{N}\sum_{b_{2}\neq b_{1}}^{N}\me^{-\mi(\mathbf{k}_{d}-\mathbf{k}_{w}).\mathbf{r}_{b_{1}}}\me^{\mi(\mathbf{k}_{d}-\mathbf{k}_{w}).\mathbf{r}_{b_{2}}},\\
    &\langle \hat{Z}^{2}\rangle= N+n(n-1)-2n(N-n)v^{z}+\left(1-\delta_{2,N}\right)\left(1-\delta_{n,N-1}\right)(N-n)(N-(n+1))(v^{z})^{2}.
\end{align}
with $v^{z}=-\cos(\bar{\theta})$. Taking $\mathbf{k}_{d}=\mathbf{k}_{w}$, $N>2$, $1\leq n <N-1$ and $\xi^{2}=((N-1)(\Delta \hat{Z})^{2}+\langle\hat{Z}^{2}\rangle)/(\langle\hat{X}_{\mathbf{k}_{w}}^{2}\rangle+\langle\hat{Y}_{\mathbf{k}_{w}}^{2}\rangle+\langle\hat{Z}^{2}\rangle-2N)$, we get
\begin{align}
    \xi^{2}= \frac{N^{2}+(n^{2}-n N)(1-\cos(\bar{\theta}))^{2}}{N+(N-1)N\cos^{2}(\bar{\theta})-(n^{2}-n(2N-1))\sin^{2}(\bar{\theta})}.
    \label{eq:remark3}
\end{align}
Taking the condition $\xi^{2}<1$, we get:
\begin{align}
    -2n+N(1+\cos(\bar{\theta}))<1+\cos(\bar{\theta}) \xrightarrow{} n>(N-1)\cos^{2}(\bar{\theta}/2).
\end{align}
\section{General spin squeezing parameters and connection with Kitagawa and Ueda's.}

In this section, we derive the spin squeezing parameter used in the main text to study entanglement. The quadratures and population operator are given by
\begin{align}
&\hat{X}_{\mathbf{k}_{w}}=\sum_{p=1}^{N}\hat{\sigma}_{\mathbf{k}_{w},p}^{x},\ \hat{Y}_{\mathbf{k}_{w}}=\sum_{p=1}^{N}\hat{\sigma}_{\mathbf{k}_{w},p}^{y},\ \hat{Z}=\sum_{p=1}^{N}\hat{\sigma}_{p}^{z}.
\end{align}
Taking into account an arbitrary direction $\hat{n}$, we define the general operator
\begin{align}
    \hat{E}_{\hat{n}}&=(\hat{X}_{\mathbf{k}_{w}}\hat{x}+\hat{Y}_{\mathbf{k}_{w}}\hat{y}+\hat{Z}\hat{z})\cdot \hat{n},\\
    &=\sum_{p=1}^{N}\hat{\sigma}^{\hat{n}}_{p}.
\end{align}
where $\hat{\sigma}^{\hat{n}}_{p}=(\hat{\sigma}_{\mathbf{k}_{w},p}^{x}\hat{x}+\hat{\sigma}_{\mathbf{k}_{\xi},p}^{y}\hat{y}+\hat{\sigma}^{z}_{p}\hat{z})\cdot\hat{n}$. Now defining two orthogonal directions $(\hat{n}_{1},\hat{n}_{2})$ and using the fact that $\hat{\sigma}_{\mathbf{k}_{w},p}^{x}\hat{\sigma}_{\mathbf{k}_{w},p}^{y}=\mi\hat{\sigma}^{z}_{p}$, $\hat{\sigma}^{z}_{p}\hat{\sigma}_{\mathbf{k}_{w},p}^{x}=\mi\hat{\sigma}_{\mathbf{k}_{w},p}^{y}$, $\hat{\sigma}_{\mathbf{k}_{w},p}^{y}\hat{\sigma}^{z}_{p}=\mi\hat{\sigma}_{\mathbf{k}_{\xi},p}^{x}$ we obtain
\begin{align}
    \nonumber[\hat{\sigma}^{\hat{n}_{1}}_{p},\hat{\sigma}^{\hat{n}_{2}}_{p}]&=2\mi(\hat{y}\cdot\hat{n}_{1}\hat{z}\cdot\hat{n}_{2}-\hat{y}\cdot\hat{n}_{2}\hat{z}\cdot\hat{n}_{1})\hat{\sigma}_{\mathbf{k}_{w},p}^{x}+2\mi(\hat{x}\cdot\hat{n}_{2}\hat{z}\cdot\hat{n}_{1}-\hat{x}\cdot \hat{n}_{1}\hat{z}\cdot\hat{n}_{2})\hat{\sigma}_{\mathbf{k}_{w},p}^{y}+2\mi(\hat{x}\cdot\hat{n}_{1}\hat{y}\cdot\hat{n}_{2}-\hat{x}\cdot\hat{n}_{2}\hat{y}\cdot\hat{n}_{1})\hat{\sigma}_{p}^{z},\\
    &=2\mi\hat{\sigma}^{\hat{n}_{3}}_{p}.
\end{align}
where we have defined the vector $\hat{n}_{3}=(\hat{y}\cdot\hat{n}_{1}\hat{z}\cdot\hat{n}_{2}-\hat{y}\cdot\hat{n}_{2}\hat{z}\cdot\hat{n}_{1},\hat{x}\cdot\hat{n}_{2}\hat{z}\cdot\hat{n}_{1}-\hat{x}\cdot \hat{n}_{1}\hat{z}\cdot\hat{n}_{2},\hat{x}\cdot\hat{n}_{1}\hat{y}\cdot\hat{n}_{2}-\hat{x}\cdot\hat{n}_{2}\hat{y}\cdot\hat{n}_{1})$, which satisfies $\hat{n}_{1}\cdot \hat{n}_{3}=0$ and $\hat{n}_{2}\cdot \hat{n}_{3}=0$. It implies that the operators $\{\hat{\sigma}^{\hat{n}_{1}}_{p},\hat{\sigma}^{\hat{n}_{2}}_{p},\hat{\sigma}^{\hat{n}_{3}}_{p}\}$ define a proper basis. Therefore, taking the triplet $\{\hat{n}_{1},\hat{n}_{2},\hat{n}_{3}\}$, the electric field-bases entanglement witnesses inequalities will read \cite{Rosario_2024}
\begin{align}
& w_{1,\mathbf{k}} = (\Delta \hat{E}_{\hat{n}_{1}})^{2}+(\Delta \hat{E}_{\hat{n}_{2}})^{2}+(\Delta \hat{E}_{\hat{n}_{3}})^{2}  -2N,\\
   & w_{2,\mathbf{k}} = 2N +(N-1)(\Delta \hat{E}_{\hat{n}_{1}})^{2}-\langle \hat{E}_{\hat{n}_{2}}^{2}\rangle -\langle \hat{E}_{\hat{n}_{3}}^{2} \rangle,\\  &w_{3,\mathbf{k}} =(N-1)\left[(\Delta \hat{E}_{\hat{n}_{1}})^{2}+(\Delta \hat{E}_{\hat{n}_{2}})^{2}\right]-\langle \hat{E}_{\hat{n}_{3}}^{2} \rangle -N (N-2).
\end{align}
The previous quantities can be rewritten as spin squeezing parameters. Introducing the quantity $\hat{E}^{2}=(\hat{E}^{\hat{n}_{1}})^{2}+(\hat{E}^{\hat{n}_{2}})^{2}+(\hat{E}^{\hat{n}_{3}})^{2}$, we get the spin squeezing parameters
\begin{align}
&\xi^{2}_{1}=\frac{(\Delta \hat{E}_{\hat{n}_{1}})^{2}+(\Delta \hat{E}_{\hat{n}_{2}})^{2}+(\Delta \hat{E}_{\hat{n}_{3}})^{2}}{2N},\\
    &\xi^{2}_{2}=\frac{\text{min}_{\hat{n}_{1}}\left((N-1)(\Delta \hat{E}_{\hat{n}_{1}} )^{2}+\langle (\hat{E}_{\hat{n}_{1}} )^{2}\rangle\right)}{\langle \hat{E}^{2}\rangle - 2N},\\
    &\xi^{2}_{3}=\frac{(N-1)\left[(\Delta \hat{E}_{\hat{n}_{1}})^{2}+(\Delta \hat{E}_{\hat{n}_{2}})^{2}\right]}{\langle \hat{E}_{\hat{n}_{3}}^{2} \rangle + N (N-2)}.
\end{align}
with the Kitagawa and Ueda parameter being a special case of $\xi^{2}_{2}$, which has been used in the main text.

\section{Minimum in the Large-$N$ Limit}
Let the factor for the quantum state in Eq.\eqref{eq:driven_state} be:
 \begin{align}
     \xi^{2} =\frac{\nu^{2}+N(N-\nu)}{N+\nu(\nu-1)+\frac{2\nu(N-\nu)}{N(N-1)}f(\mathbf{k}_{d}-\mathbf{k}_{w})},
     \label{xi_remark_2_app}
 \end{align}
considering $f(\mathbf{k}_{d}-\mathbf{k}_{w})=N(N-1)$ and $\nu$ as a continuous variable, we get the minimum of $\xi^{2}$ as $\partial(\xi^{2})/\partial \nu =0$. Then, from the derivative condition we obtain that the minimum of $\xi^{2}$ for arbitrary $N$ is given by
\begin{align}
    \nu=\frac{-N-N^{2}+\sqrt{N^{2}+3N^{4}}}{N-1}.
\end{align}
Note that for large $N$, the latter expression converges to $\nu\approx (\sqrt{3}-1)N$. Now, using this value in Eq.\eqref{xi_remark_2_app} when $N\xrightarrow{} \infty$ we get:
\begin{align}
    \xi^{2}=\frac{3(2-\sqrt{3})}{4\sqrt{3}-6}=\frac{\sqrt{3}}{2}.
\end{align}

\end{document}